\documentclass{IEEEtran}

\usepackage{amssymb,amsmath,color,graphicx, amsbsy}
\usepackage{graphicx}
\usepackage{subfigure}
\usepackage{cite}
\usepackage{algorithm}
\usepackage{algorithmic}
\usepackage{amsmath}
\usepackage{amsfonts}
\usepackage{amssymb}
\usepackage{graphicx}
\usepackage{epstopdf}
\usepackage{lettrine}

\begin{document}

\title{A Data-Driven Study to Discover, Characterize, and Classify Convergence Bidding Strategies in California ISO Energy Market}

%\title{\fontsize{22.5pt}{27pt}\selectfont{A Data-Driven Study to Discover, Characterize, and Classify Convergence Bidding Strategies in California ISO Energy Market}}

\author{Ehsan Samani,~\IEEEmembership{Student Member, IEEE}
and Hamed Mohsenian-Rad,~\IEEEmembership{Fellow, IEEE\vspace{-0.05cm}}\\
Department of Electrical and Computer Engineering, University of California, Riverside, CA\\
e-mails: ealia001@ucr.edu and hamed@ece.ucr.edu
\thanks{The authors are with the Department of Electrical and Computer Engineering, University of California, Riverside, CA, USA, 92521. This work was supported by the National Science Foundation (NSF) grant 1711944. The corresponding author is Hamed Mohsenian-Rad; e-mail: hamed@ece.ucr.edu.}\vspace{-0.5cm}}

\maketitle

%%%%%%%%%%%%%%%%%%%%%%%%%%%%%%%%%%%%%%%%%%%%%%%%%%%%%%%%%%%%%%%%%%%
\thispagestyle{empty}

\pagestyle{empty}

\begin{abstract}

Convergence bidding has been adopted in recent years by most Independent System Operators (ISOs) in the United States as a relatively new market mechanism to enhance market efficiency. Convergence bidding affects many aspects of the operation of the electricity markets and there is currently a gap in the literature on understanding how the market participants strategically select their convergence bids in practice. To address this open problem, in this paper, we study three years of real-world market data from the California ISO energy market. 
%First, a data-driven overview of all submitted convergence bids (CBs) is provided; and the performance of each individual convergence bidder is studied based on the number of their submitted CBs, the number of locations that they placed the CBs, percentage of submitted supply or demand CBs, amount of cleared CBs, and their gained profit or loss.
%
First, we provide a data-driven overview of all submitted convergence bids (CBs) and analyze the performance of each individual convergence bidder based on the number of their submitted CBs, the number of locations that they placed the CBs, the percentage of submitted supply or demand CBs, the amount of cleared CBs, and their gained profit or loss. 
Next, we scrutinize the bidding strategies of the 13 largest market players that account for 75\% of all CBs in the California ISO market. We identify quantitative features to characterize and distinguish their different convergence bidding strategies. This analysis results in revealing three different classes of CB strategies that are used in practice. We identify the differences between these strategic bidding classes and compare their advantages and disadvantages. We also explain how some of the most active market participants are using bidding strategies that do not match any of the strategic bidding methods that currently exist in the literature.

\vspace{0.2cm}


\end{abstract}
\vspace{-0.1cm}
%%%%%%%%%%%%%%%%%%%%%%%%%%%%%%%%%%%%%%%%%%%%%%%%%%%%%%%%%%%%%%%%%%%%
\section{Introduction} \label{sec:introduction}
%%%%%%%%%%%%%%%%%%%%%%%%%%%%%%%%%%%%%%%%%%%%%%%%%%%%%%%%%%%%%%%%%%%%
%\subsection{Background} 
Convergence bidding, a.k.a., virtual bidding, is a market mechanism that is used by Independent System Operators (ISOs) in two-settlement electricity markets to reduce the price gap between the day-ahead market (DAM) and the real-time market (RTM) in order to increase market efficiency \cite{hogan2016,2015Financial0MIT}. 
\color{black}
Some of the potential advantages that are identified for supporting convergence bids (CBs) include the following: improving the efficiency of the day-ahead commitment and energy schedules, reducing the cost of hedging, allowing for efficient settlement of
financial transmission right, and making it advantageous
for parties to utilize the liquidity provided in the market \cite{hogan2006revenue}.
%
%ISOs expect convergence bidders to provide highly competitive
%pressure to arbitrage between the markets and achieve price
%convergence.
%
CBs can also affect the integration of renewable energy resources and demand response resources in electricity markets, e.g., see \cite{kazempour2017value, woo2015virtual}.
Several ISOs in the United States, including the California ISO, currently use CBs \cite{Ref_CAISO_CB_Document}.

While the basic principles of convergence bidding are studied in the academic literature and also industry reports, there is currently a gap in this field about understanding the strategy and behavior of market participants that submit CBs in practice; such as in the California ISO and elsewhere. %This is a critical subject because the way that market participants select their CBs can ultimately shape the impact of CBs on electricity markets.  
Addressing this open problem is the focus of this paper.

\subsection{Summary of Discoveries and Contributions} \label{sec:intr:cntr}
To the best of our knowledge, this paper is the first study on real-world convergence bidding strategies in the California ISO energy market; or other ISO markets. The main discoveries and contributions in this paper are  as follow:
\begin{itemize}
\vspace{0.1cm}
\item Every CB that is submitted to the California ISO energy market over the past three years is analyzed and the results are summarized in terms of the type and the number of the bids, the number of participated market locations, the amount of cleared CBs, and the amount of profits gained (or losses) by CB market participants.

\vspace{0.1cm}

\item Based on the characteristics and data-driven features of the analyzed CBs, we identify three different classes of bidding strategies that are currently being used in practice. Drastic differences between these classes of convergence bidding strategies are identified and some of their advantages and disadvantages are investigated.

\vspace{0.1cm}

\item Our analysis explains how some of the most active CB market participants are currently using some interesting bidding strategies that do \emph{not} match any of the strategic bidding methods that currently exist in the literature. Some of them have  changed their strategies over time. The results in scrutinizing the bidding strategies of real-world market participants can help ISOs with their ongoing efforts to improve the efficiency in electricity markets.

%better understanding \color{black} \# [Hamed: please complete this sentence properly.] \color{black}

\vspace{0.1cm}
\end{itemize}
\color{black}

%\cite{samani2020cb}. 

%At each pair of location and time, when virtual bidders predict a higher (lower) price in DAM than RTM, they should offer (bids) to sell (buy) a specific amount of energy in DAM and buy (sell) the same amount of energy in the RTM.
% \vspace{-0.1cm}
 \subsection{Literature Review} \label{sec:intr:lit}
% \color{black} Despite the fact that CBs are widely adopted by ISOs in recent years, there is still limited literature on addressing the issues related to CBs in electricity markets. \color{black}

The literature on CBs can be broadly divided into two groups. First, there are  studies that \emph{assess the impact} of CBs on electricity markets
%\color{black}
\cite{2016Dynamic0Theis,oren2015,nyc2007,birge2018limits,tang2016model,hadsell2007one,wolak2013,samani2020cb}.
The impact of CBs on market efficiency are studied in \cite{oren2015}. In \cite{samani2020cb}, a methodology is proposed to identify under what theoretical conditions a CB results in price divergence, instead of price convergence. The analysis is done in nodal electricity markets and factors such as \emph{transmission line congestion} are investigated.
\color{black}
%
%. Also, 
%
%market efficiency before and after using CBs are compared in these papers.
%
%
%
%CBs have been studied in the literature for two different purposes. First, understanding their impact on the market \cite{oren2015,wolak2013}. The market efficiency before and after using CBs are compared in these papers. In
%in \cite{samani2020cb},
%structural characteristics of CBs in the nodal electricity market are analyzed to find the situations in which divergence happens in the market. 
%

Second, there are studies that \emph{develop strategies} for convergence bidders to maximize their profit \cite{baltaoglu2018algorithmic,xiao2018risk}.
In \cite{baltaoglu2018algorithmic}, an online learning algorithm is proposed to maximize the cumulative payoff over a finite number of trading sessions. In \cite{xiao2018risk},  a stochastic optimization model is proposed to optimally place CBs under different risk management considerations; the scenarios of electricity prices are generated by using the seasonal autoregressive integrated moving average model.

%the impact of CBs on market efficiency and on price convergence are studied in \cite{oren2015} and \cite{samani2020cb}, respectively. 

%strategies to submit CBs \cite{baltaoglu2018algorithmic}, \cite{xiao2018risk}. Basically all the proposed methods for optimal bidding are based on forecasting the locational marginal prices (LMPs). 

%In this paper, for the first time, the performance and strategy of each individual participant is analyzed. 
\color{black}
The study in this paper is more aligned with the second group of papers mentioned above. However, here, we do \emph{not} propose any new theoretical strategy for convergence bidding. Instead, we do a data-driven study based on real-world market data to \emph{discover}, \emph{characterize}, and \emph{classify} the strategies that have been used by market participants in practice over the past three years in the California ISO energy market.

\begin{figure}[t]
 \centering
{\scalebox{0.62}{\includegraphics*{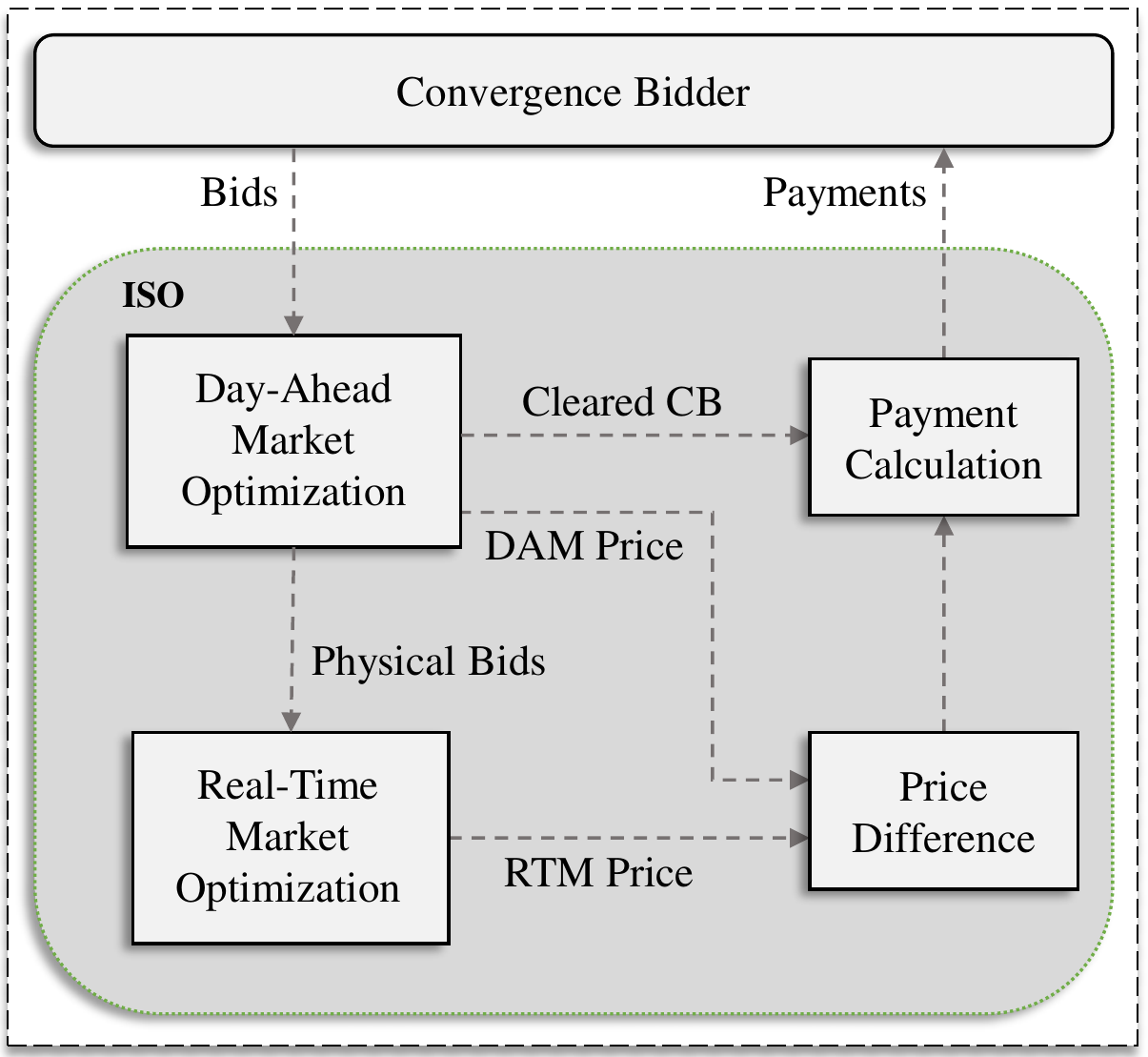}}}
\vspace{0.05cm}
\caption{Overview of the convergence bidding process in an ISO market \cite{samani2020cb}.}
\label{fig:out1}
\end{figure}

%can be summarized as follows:
%\begin{enumerate}
%  \item Each individual convergence bidder in California ISO energy market is analyzed and their percentage of participation in the market, number of participated locations, percentage of submitted supply or demand bids, amount of cleared energy, and amount of profit or loss are investigated. 
%  \vspace{0.1cm}
  
%  \item Based on two introduced features for individual convergence bidders, percentage of being awarded in the market, and the distance of submitted bid prices from LMPs, three different classes of bidding strategies that are currently being used in practice are introduced. Drastic differences between these different strategic bidding classes are identified and some of their advantages and disadvantages are explored. 

 %  \vspace{0.1cm}
  
%  \item  Our Analysis explain how some of the most active participants are using bidding strategies that do not match any of the strategic bidding methods that currently exist in the literature.
%\end{enumerate}

\section{Overview of the Real-World \\ California ISO Market Data} \label{sec:gnral}

\subsection{Background}

A CB can be a supply
%\color{black} offer \color{black} 
or a demand bid.
However, a CB is \emph{not} a physical bid; it is purely financial.
\color{black}
A supply (demand)\footnote{\color{black}Parenthesis is used to separate two different cases for the supply and demand CBs. The words in parentheses should be used for demand cases.} CB is a bid to sell (buy) energy in DAM and buy (sell) the \emph{same amount} of energy in RTM without any obligation to produce (consume) energy \cite{AnotherStrategicBidding, kazempour2017value,ftr2013, goodbad2010,samani2019tri,AnotherFTR}. 
If a supply (demand) CB is cleared in DAM, then the bidder is credited (charged) at the DAM price and charged (credited) at the RTM price. Thus, the difference between the earning in DAM (RTM) and the cost in RTM (DAM) will be paid to the bidder. \color{black}
The process of clearing CBs  and the related payment is outlined in Fig. \ref{fig:out1}. The payment is calculated by multiplying the cleared amount of energy by the difference between DAM and RTM prices. 

\subsection{Data Set}

\begin{figure}[t]
 \centering
{\scalebox{0.7}{\includegraphics*{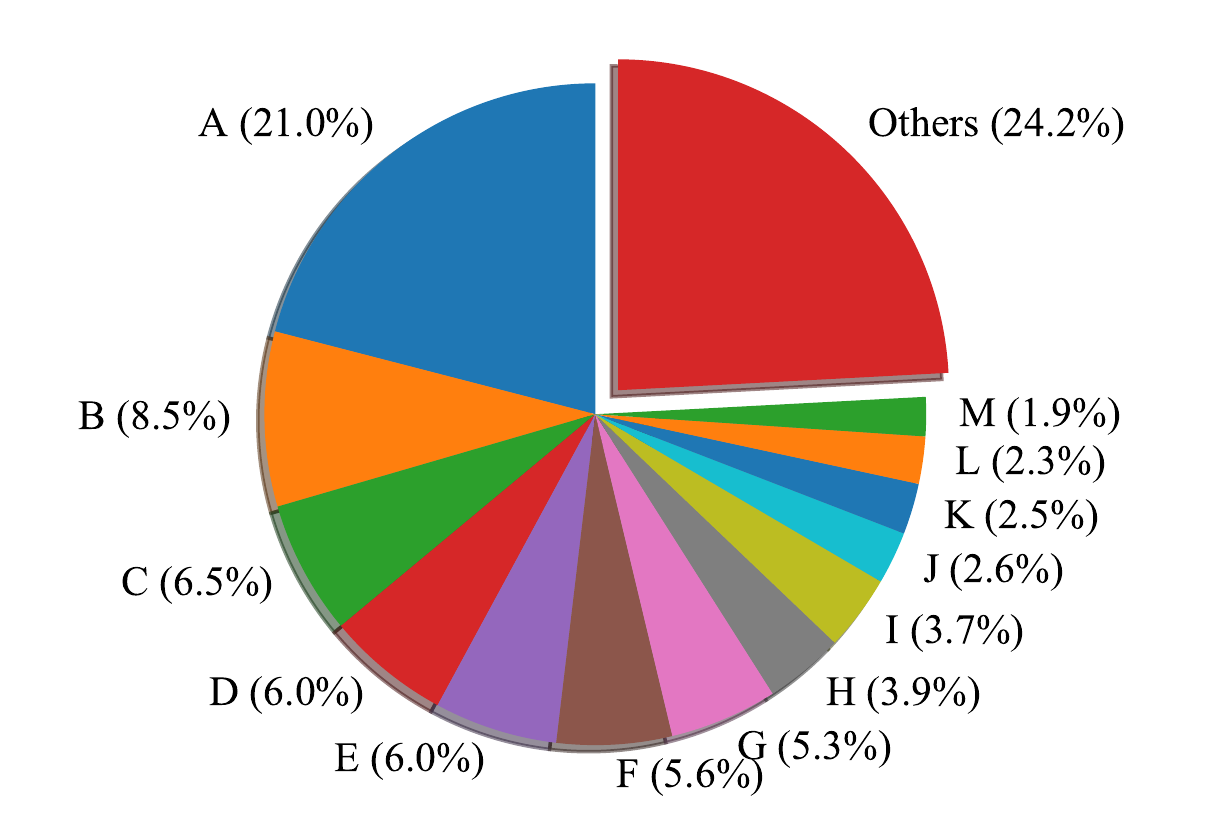}}}
\caption{Share of the CB market across a total of 101 convergence bidders. The Alias IDs A to M are marked for the 13 largest market participants.}
\label{fig:prcnt}
\end{figure}

Three years (2017-2019) of data from the California ISO energy market are examined, including the size, price, and location of every submitted CB, day-ahead locational marginal prices (D-LMPs), real-time LMPs (R-LMPs), and the net cleared CB awards. %, and convergence bidding public bids are analyzed in this paper. We matched the public bids data to the other three data sets in order to analyze each individual market participant. %In this section, first, the basic observations of market related to convergence bidding are provided and then the performance of each individual participant is investigated.
%It should be noted that the 
Focusing on the aggregate pricing nodes (APnodes) in the California ISO energy market, a total of 2265 APnodes are examined; out of which 475 APnodes hosted at least one CB during the three-years period of this study. Every month, on average, a total of 387 APnodes hosted any CB.  

\begin{figure}[t]
 \centering
 \vspace{0.4cm}
{\scalebox{0.37}{\includegraphics*{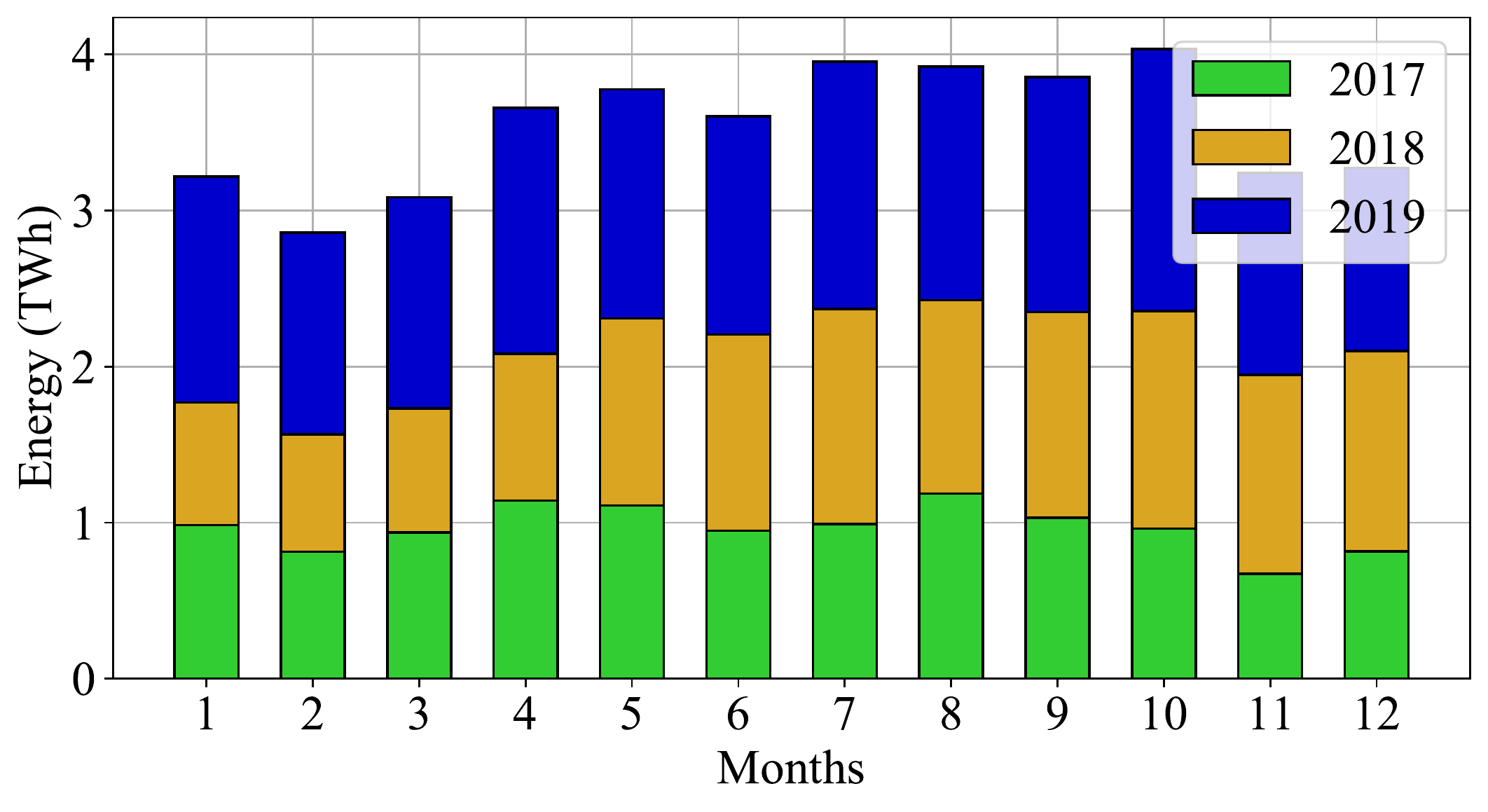}}} \vspace{-0.05cm}
\caption{Total monthly amount of cleared energy by CBs in each year.}
\label{fig:mon_energy}
\end{figure}

\subsection{Total Bids and Total Profit}

The total number of market participants that ever submitted a CB during the three-years period of this study was 101, with a monthly average of 52 market participants. A total of 13 market participants dominated the market. Their market shares are shown in Fig. \ref{fig:prcnt}; where we assigned them \emph{Alias IDs}, denoted by A to M. These 13 market participants accounted for over 75\% of all the CBs that were submitted in this period. 

\begin{figure}[t]
 \centering
{\scalebox{0.37}{\includegraphics*{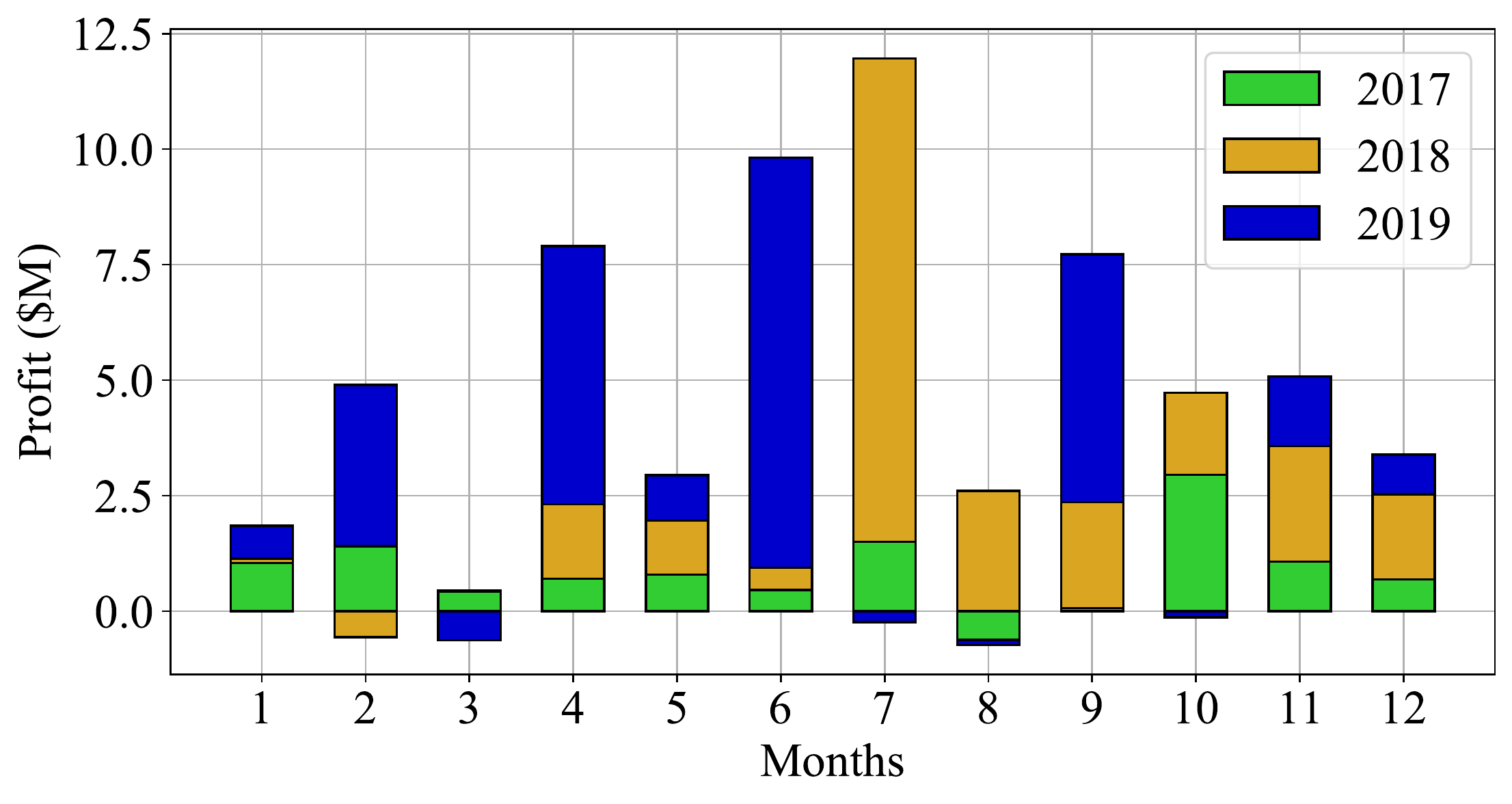}}}
\caption{Total monthly net profit by convergence bidders at each year.}
\label{fig:mon_prof}
\end{figure}

The total profit that was earned by all the market participants in the CB market during this period was \$61 Million. Out of the 101 convergence bidders, 74 of them made money, i.e., had a net positive profit.  Fig. \ref{fig:mon_energy} shows the total monthly amount of cleared energy at each year for the convergence bidders and Fig. \ref{fig:mon_prof} shows the net monthly profit that all the convergence bidders earned during this period of study. \color{black} 
Interesting, there were months were the market participants had an overall loss, i.e., negative profit as the outcome of their convergence bidding. Another interesting observation is that even though the net profit fluctuated significantly across different months; the amount of cleared CB was about the same in each month. %Profit or loss is equal to the cleared amount of energy multiply by the difference between D-LMP and R-LMP at each pair of location time.
%The question is what was the difference of each bidder's strategy which results in a wide range of profit. For both participants and ISOs, the amount of money that each convergence bidder earns in the market is the key characteristic of their performance.

\vspace{0.2cm}
\section{Analysis of Individual Convergence \\ Bidders and Their Bidding Strategies}
\label{sec:strtg}

In this section, we focus on the 13  dominant market participants and scrutinize their bidding performance and strategies.

\vspace{-0.2cm}
\subsection{Performance of Individual Convergence Bidders} \label{sec:gnral:attr}
%Other than the percentage of participation in the market, four other unique characteristic of each individual bidder showing their performance in the market are analyzed. 
Table \ref{tab1} shows the average monthly amount of cleared CB, the average monthly amount of profit, the number of participated locations, and the percentage of submitted supply CBs for each of the 13 dominant market participants.

Some of the convergence bidders placed CBs in almost all the locations; such as Alias ID D that placed CBs in 461 (out of 475) APnodes. Some others placed CBs in only a  few locations; such as Alias ID M that placed CBs in 21 APnodes. 

Most of these 13 dominant convergence bidders submitted both supply and demand bids. But some of them, such as Alias ID B, submitted one type of bid more than the other type.

The highest profit was earned by Alias ID I at about \$290K per month. This convergence bidder had only 3.7\% share of the market. As Alias ID A that had the highest share of the market at 21.0\%, it only earned about \$55K per month.

\color{black} Based on the above results, we can see that increasing the number of locations for placing CBs does \emph{not} necessarily result in increasing profit.
There is no direct relationship between the number of nodes that a market participant submits CBs and their total profit.
Alias ID I with the highest profit, placed CBs in only 88 APnodes, which is the second lowest in the Table. Also, the convergence bidder with the lowest profit, with Alias ID K, bids in 326 APnodes which is among the highest.
%Also, alias ID A as the most active convergence bidder in term of number of total submitted CBs, has submitted CB in about two times nodes compare to Alias ID I while their profit was about one fifth of Alias ID I.
By comparing Alias ID A as the most active convergence bidder in term of the number of submitted CBs and Alias ID I as the most lucrative market participant, we see Alias ID I participated in about half of the number of nodes compared to Alias ID A and earned about five times more profit. \color{black}

Last but not least, as we see in Table I, the range of average monthly profit for each of these 13 dominant convergence bidders is very wide, from about \$2K to \$290K.

\vspace{-0.2cm}
\subsection{Classification of Bidding Strategies}
We introduce two quantitative features to %analyze the strategies that convergence bidders use in the California ISO energy market, two features are introduced to %
characterize the CBs that are submitted to the California ISO energy market:
1) The percentage of the bids being awarded in the market;
2) The average distance of submitted price bids from D-LMPs.

% \begin{figure}[t]
%  \centering
% {\scalebox{0.41}{\includegraphics*{P2.pdf}}}
% \caption{Distance from LMP, i.e., $\Delta$: a) in a supply bid; b) in a demand bid.}
% \label{fig:dist_def}
% \end{figure}

\begin{figure}[h]
 \centering
{\scalebox{0.42}{\includegraphics*{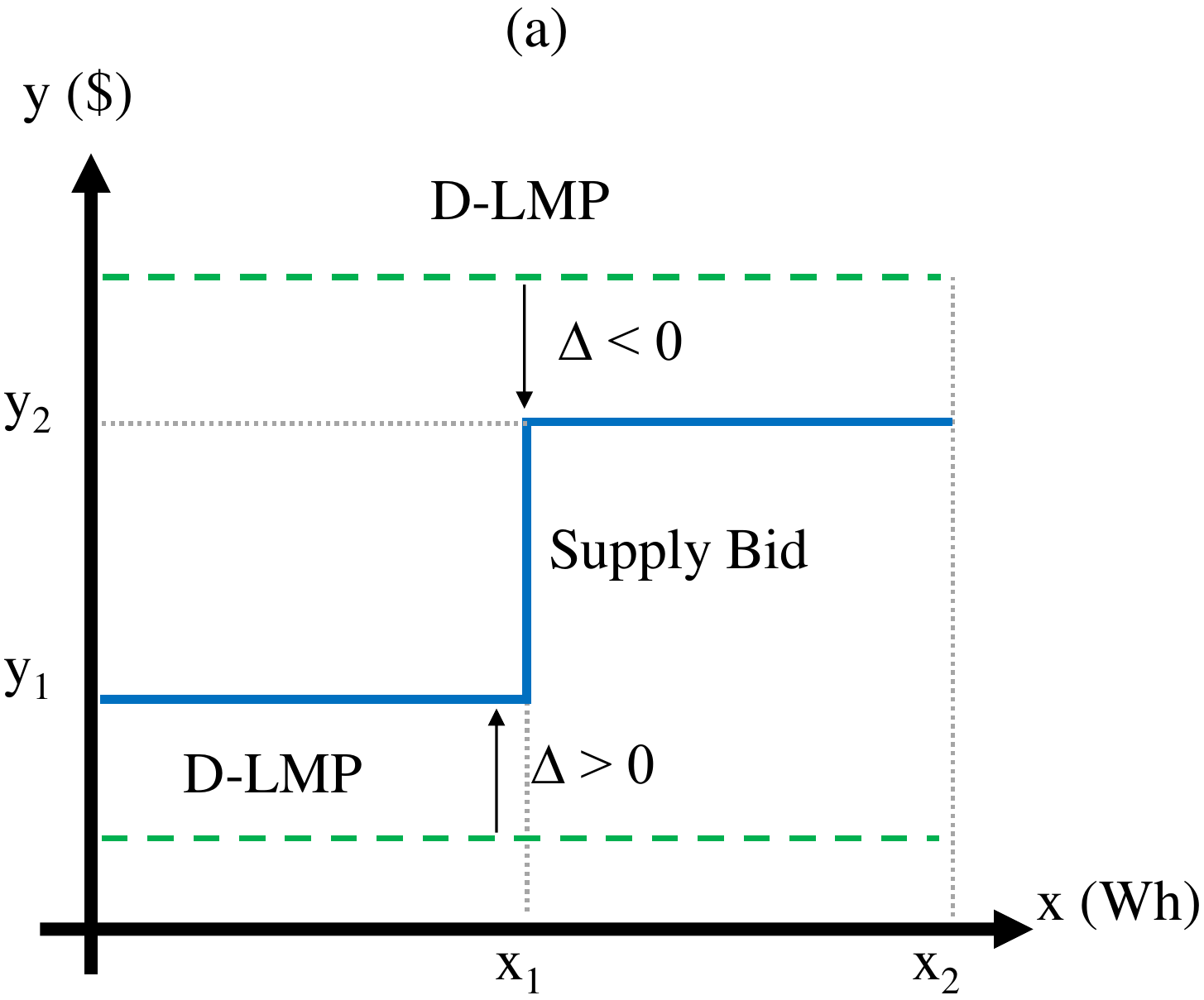}}}
% \label{fig:dist_}
{\scalebox{0.42}{\includegraphics*{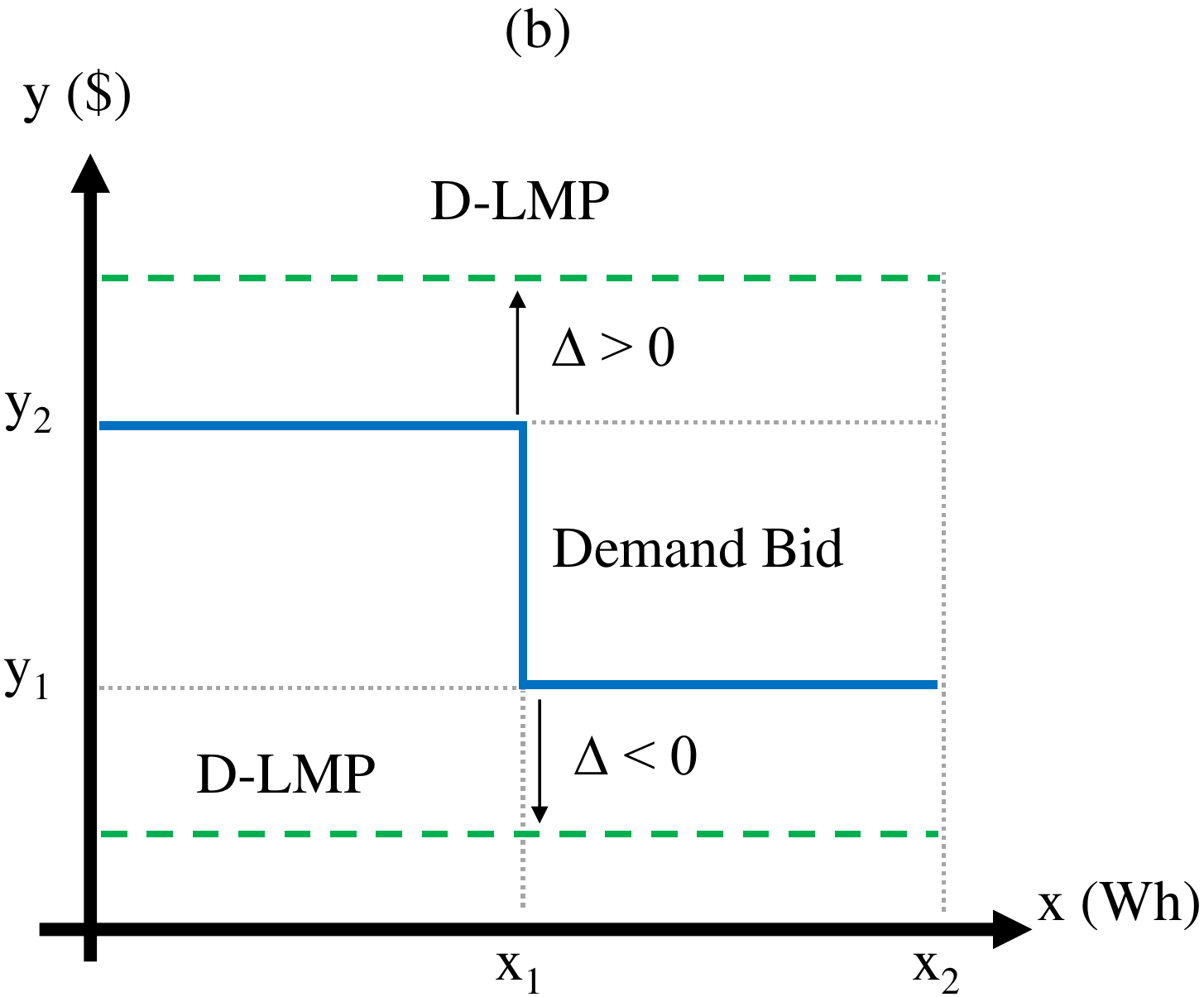}}}
\caption{Distance from LMP, i.e., $\Delta$: a) in a supply bid; b) in a demand bid.}
\label{fig:dist_def}
\end{figure}

The second feature needs some explanation. 
%
%Percentage of being awarded in the market and the average distance of submitted price bids from D-LMPs.  
Fig. \ref{fig:dist_def} shows the definition of \emph{distance from LMP}, denoted by $\Delta$,  for piece-wise linear supply and demand CBs. If $\Delta > 0$ for a supply (demand) CB then the price component of the CB is  \emph{higher} (\emph{lower}) than the D-LMP. CBs with  $\Delta > 0$ are \emph{not} cleared by the ISO while CBs with $\Delta \leq 0$ are cleared in the market. %by the ISO. 

\begin{figure}[t]
 \centering
{\scalebox{0.38}{\includegraphics*{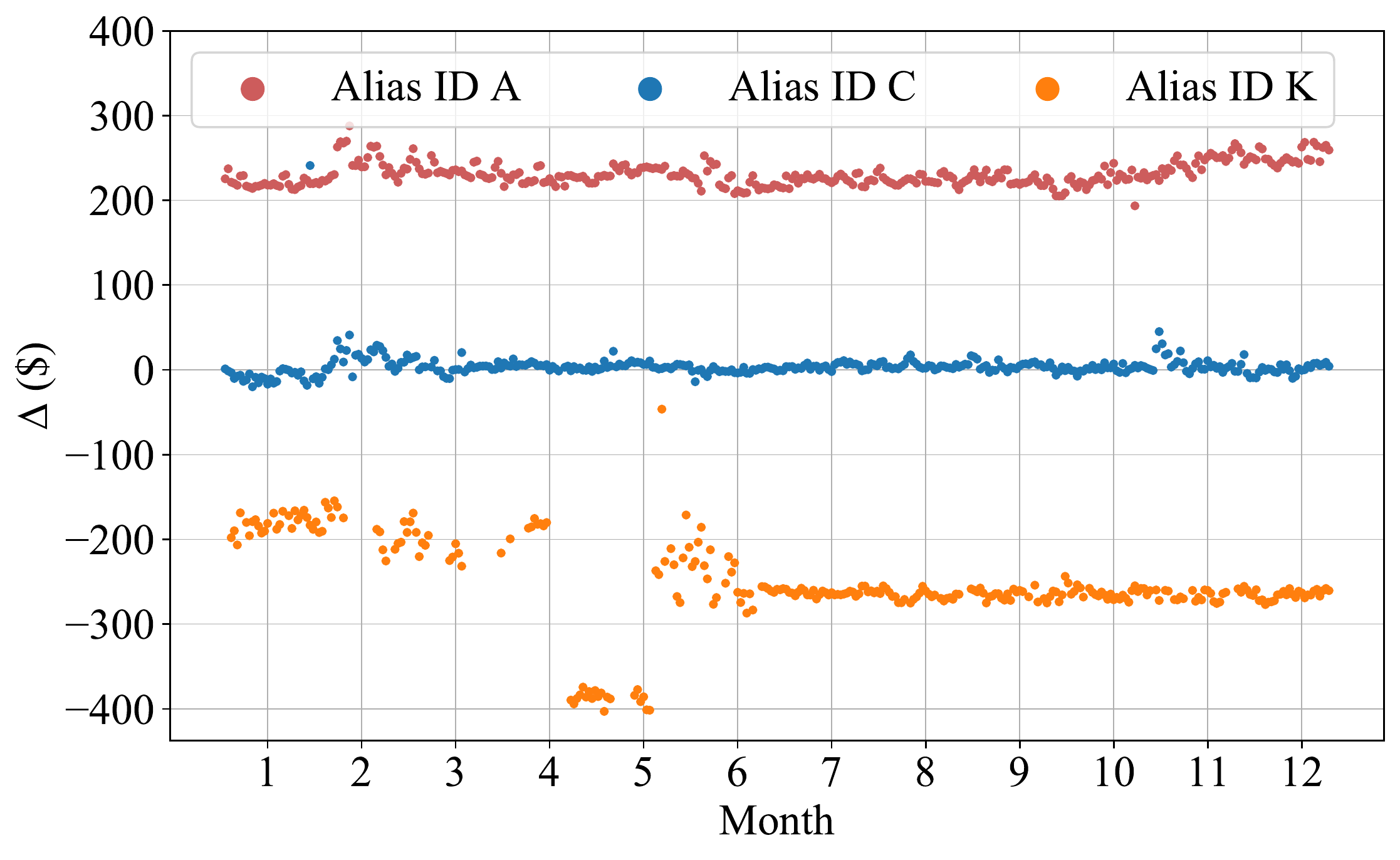}}}
\vspace{-0.1cm}
\caption{The average daily distance of the price component of the submitted CBs in the market from D-LMPs for three representative market participants.}
\label{fig:Dist_2019}
\end{figure}

\begin{figure}[t]
 \centering
{\scalebox{0.38}{\includegraphics*{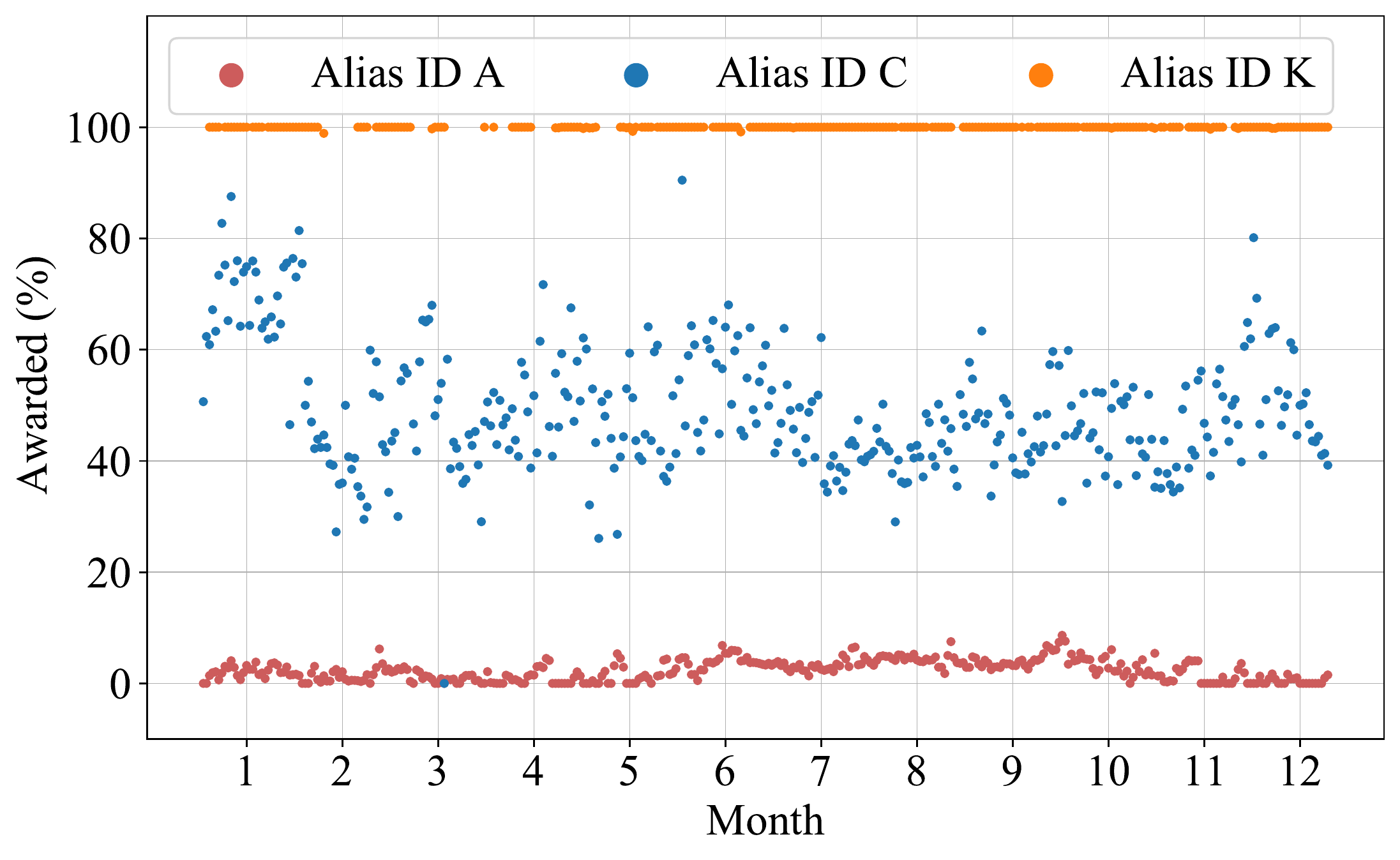}}}
\vspace{-0.1cm}
\caption{The average daily percentage of CBs that are being awarded in the market for three representative market participants.}
\label{fig:Dist_2019_aw}
\end{figure}

 Fig. \ref{fig:Dist_2019} shows the average daily distance of the price component of the submitted convergence bids in the market from D-LMPs for three representative market participants during one year in 2019. Also, Fig. \ref{fig:Dist_2019_aw} shows the average daily percentage of being awarded in the market for them. \color{black}
There are \emph{clear distinctions} among these three market participants in terms of both features. CBs from Alias ID A have \emph{large positive} distance from LMPs, i.e., $\Delta \gg 0$; as a result; they are almost always \emph{not} awarded by the ISO. CBs from Alias ID K have \emph{large negative} distance from LMPs, i.e., $\Delta \ll 0$; as a result; they are almost always \emph{awarded} by the ISO. CBs from Alias ID C have \emph{small} positive or negative 
distance from LMPs, i.e., $\Delta \approx 0$; and they are \emph{usually} awarded by the ISO. If we plot these two features for the rest of the 13 dominant market participants, they  all follow one of these three scenarios. 

%As we can see, these three convergence bidders have very distinctive features; as three colors of dots are very clearly separate. \color{black}

% \begin{figure}[t]
%  \centering
% {\scalebox{0.38}{\includegraphics*{Selected_2019.pdf}}}
% \label{fig:dist_}
% {\scalebox{0.38}{\includegraphics*{Dist_2019.pdf}}}
% \caption{Classification features for three representative market participants: (a) the daily percentage of being awarded in the market; (b) average daily $\Delta$.}
% %\caption{Comparing the features of the three CB strategies based on a representative market participant for each strategy: (a) the \textcolor{black}{average monthly $\Delta$}; (b) the monthly percentage of their bid being awarded in the market.}   %for the distance of submitted price bids from D-LMPs.}
% \label{fig:perct_}
% \end{figure}

\begin {table}[t]
\centering
\caption {Bidding Performance of the  Dominant Market Participants}
 \label{tab1}
\begin{center}
   \begin{tabular}{| c | c | c | c | c |}
   \hline
Alias ID & Energy (GWh) & Profit (\$)   & Nodes & Supply CB (\%) \\ \hline
A     & 23.2                  &  55,208   & 158   & 43\%        \\ \hline
B     & 86.0                 &  94,857   & 279   & 79\%        \\ \hline 
C     & 31.2                 &  72,548   & 341   & 57\%        \\ \hline
D     & 127.7                 &  29,880     & 461   & 52\%        \\ \hline
E     & 30.4                 &  22,372     & 111   & 48\%        \\ \hline
F     & 40.0                 &  133,371   & 246   & 60\%        \\ \hline
G     & 6.1                  &  17,755     & 341   & 55\%        \\ \hline
H     & 35.9                  &  38,543     & 145   & 66\%        \\ \hline
I     & 198.6                 &  289,730  & 88    & 54\%        \\ \hline
J     & 13.9                  &  12,855     & 418   & 75\%        \\ \hline
K     & 14.4                  &  2,049      & 326   & 46\%        \\ \hline
L     & 42.9                  &  86,448   & 95    & 55\%        \\ \hline
M     & 65.6                 &  63,805   & 21    & 49\%        \\ \hline
\end{tabular}
\end{center}
\vspace{-0.4cm}
\end{table}

\subsection{Three Different Identified Bidding Strategies}

Based on the above analysis, we can distinguish three different types of CB strategies in the California ISO market: 
   
   \vspace{0.1cm}
   
\textbf{CB Strategy 1}: Forecast D-LMPs and R-LMPs at any time and location. If the foretasted D-LMP is higher (lower) than the foretasted R-LMP, then submit a supply (demand) CB. Set the price component to be lower (higher) than the foretasted D-LMP such that your supply (demand) CB is awarded; but not much lower (higher), i.e., $\Delta \approx 0$ in order to avoid entering the market when D-LMP is unexpectedly low (high) and a supply (demand) CB is submitted,
which can cause a major loss.
 
  \vspace{0.1cm}
  
\textbf{CB Strategy 2}: Set the price such that the submitted CB is almost always awarded; i.e., $\Delta \ll 0$. Use price forecasting to decide whether to submit a supply CB or a demand CB.

%even if the prices at DAM are very low (high) and a supply (demand) CB is submitted, which can result in a major loss.

%Same as CB Strategy 1 in terms of using D-LMP and R-LMP forecasts to decide on whether to submit a supply CB or a demand CB. But set the price component to be very low (high) for your supply (demand) CB such your submitted CB is almost certainly awarded by the ISO.
%
%Bid with a $\Delta \ll 0$ such that the CB is always cleared by the ISO regardless of the D-LMP. 

  \vspace{0.1cm}
  
  \color{black}

\textbf{CB Strategy 3}:  Set the price such that the submitted CB is almost always \emph{not} awarded; i.e., $\Delta \gg 0$. Importantly, 
%
%Bid with a large positive $\Delta$ such that the submitted CB is almost always \emph{not} cleared by the ISO;  but 
%
\emph{if} your supply (demand) CB happens to be cleared due to an unexpectedly high (low) DAM price, \emph{then} the profit is large. \color{black}%it goes back down (up) in the RTM and make a large profit.\color{black}%but \emph{if} the bid is cleared \emph{then} make a large profit.

  \vspace{0.1cm}

%Interestingly, while the success and profitability in CB Strategy 1 directly depends on the \emph{accuracy} of forecasting D-LMPs and R-LMPs and CB Strategies 2 needs to forecast the difference between  D-LMPs and R-LMPs to chose between a supply or demand bid, \color{black} CB Strategies 3 simply gets rid of the need to forecast D-LMPs and R-LMPs  and instead works on event analysis. \color{black}  

%This is a very important observation with respect to CB Strategies 3; because, to the best of our knowledge, there is \emph{not} currently exist any convergence bidding strategy in the electricity market literature that does not inherently require some sort of price forecasting, e.g., see \cite{baltaoglu2018algorithmic,xiao2018risk}. 

%We can explain these strategies also in terms of risk management. CB Strategy 1 is \color{black} \emph{risk-seeking} strategy but tries to lower the risk by capturing the randomness in D-LMPs and R-LMPs. CB Strategy 2 is completely a \emph{risk-seeking} strategy leaves profit entirely to the randomness in the market. \color{black} Finally, CB Strategy 3 is a \emph{risk-averse} strategy. In this strategy, it is very rare that the submitted CB is cleared; but when it is cleared \color{black} most probably it results to a large profit. \color{black}

\begin {table}
\centering
\caption {convergence bidding strategies and corresponding features.}
 \label{tab2}
\begin{center}
   \begin{tabular}{| c | c | c | c |}
   \hline
Alias ID & Awarded (\%) & Delta (\$) & Strategy  \\ \hline
A  & 3.85\%              & 211.23                 & 3                \\\hline
B  & 66.45\%             & -8.29                  & 1                \\\hline
C  & 63.20\%             & -0.10                  & 1                \\\hline
D  & 96.03\%             & -88.08                 & 2                \\\hline
E  & 11.61\%             & 122.56                 & 3                \\\hline
F  & 12.21\%             & 100.26                 & 3 to 1          \\\hline
G  & 12.14\%             & 18.04                  & 3                 \\\hline
H  & 15.56\%             & 101.19                 & 3                \\\hline
I  & 95.72\%             & -25.92                 & 1                \\\hline
J  & 58.13\%             & -6.47                  & 1                \\\hline
K  & 99.94\%             & -256.78                & 2                \\\hline
L  & 56.26\%             & -26.03                 & 1                \\\hline
M  & 77.56\%             & 2.40                   & 1                \\\hline
\end{tabular}
\vspace{-0.2cm}
\end{center}
\end{table}

Table II shows the CB Strategies that have been used by the dominant convergence bidders in the California ISO market. 
Overall, we can see that CB Strategy 1 and CB Strategy 3 have the two  popular choices. While CB strategy 1 \emph{does} generally match the strategic convergence bidding approaches in the literature  e.g., see \cite{baltaoglu2018algorithmic,xiao2018risk}; to the best of our knowledge, CB Strategy 3 has not been previously discussed in the literature. 
%
%\color{black} 
Furthermore, while Strategy 1 critically requires \emph{accurate price forecasting} to be successful, Strategy 3 does \emph{not} really need to forecast the prices and instead relies on the unexpected price fluctuations between DAM and RTM. %\color{black}

%\color{black} The presented strategies can cover all the strategies that we have seen for dominant market participants. \color{black} 

\begin{figure}[t]
 \centering
{\scalebox{0.38}{\includegraphics*{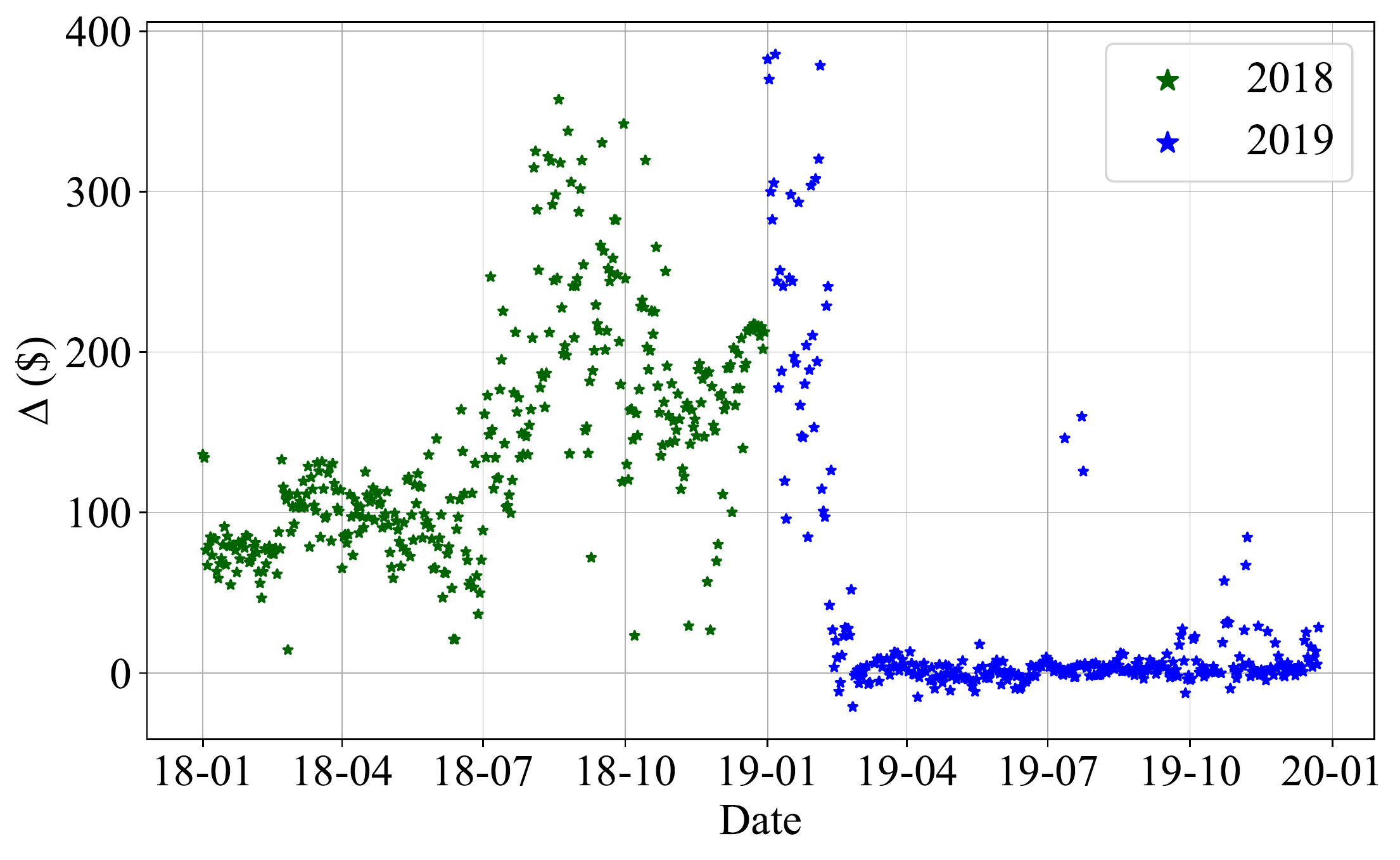}}}
\caption{The average daily distance of the price component of the submitted CBs in the market from D-LMPs for the Alias ID F. The two different colors of the points highlight the fact that the market participant changed its strategy.}
\label{fig:Dist_140124}
\end{figure}
\begin{figure}[t]
 \centering
{\scalebox{0.38}{\includegraphics*{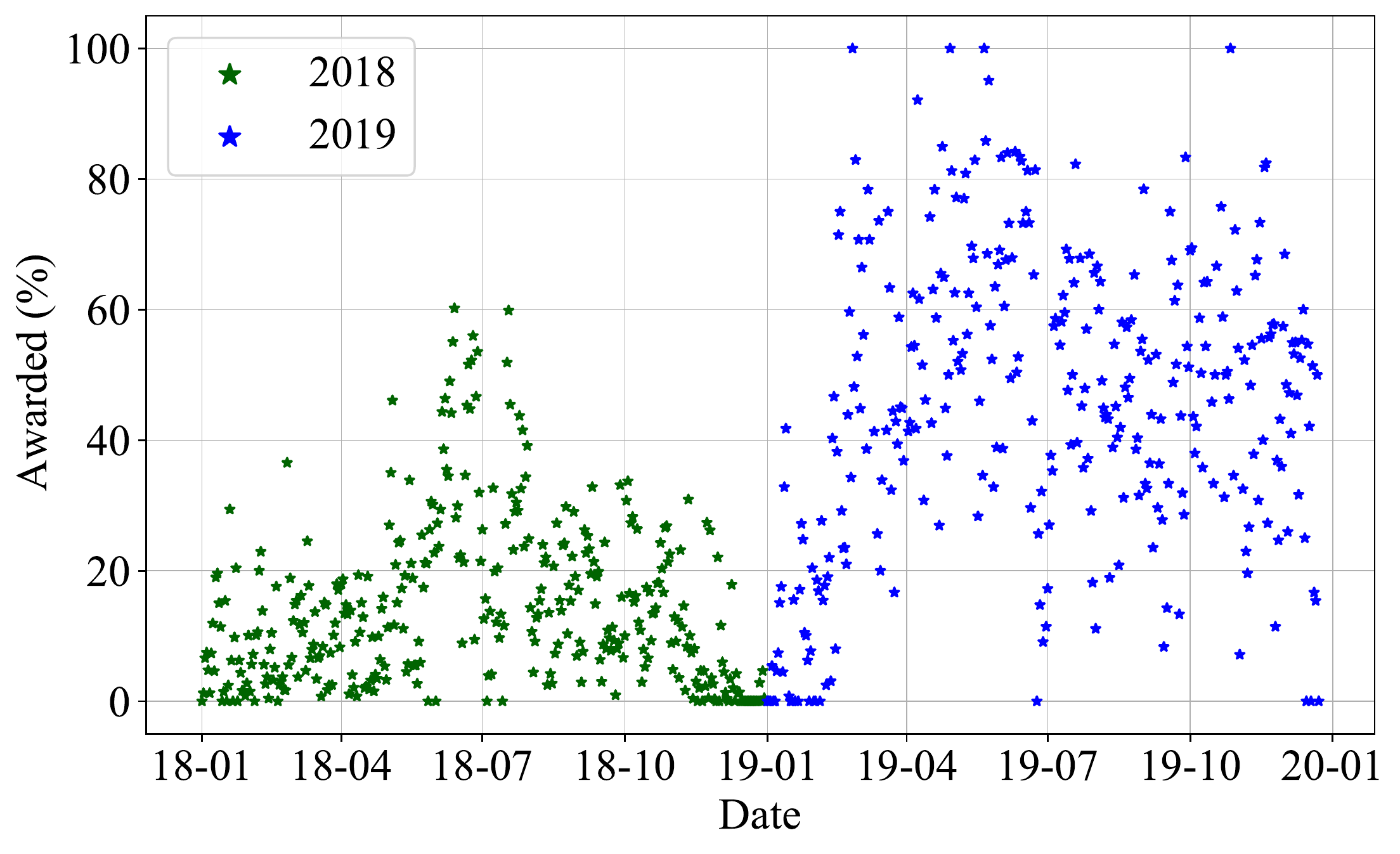}}}
\caption{The average daily percentage of being awarded in the market for Alias ID F. We can see the change in their bidding strategy.}
\label{fig:Dist_140124_aw}
\end{figure}

Regarding Alias ID F, the bidding strategy started as CB Strategy 3 but after several months switched to CB Strategy 1. Fig. \ref{fig:Dist_140124} shows the average daily distance of the price component of the submitted CBs from D-LMPs in two years (2018-2019). Fig. \ref{fig:Dist_140124_aw} shows the daily percentage of being awarded in the market. As we can see in both figures, the bidding strategy changed in early 2019 and they started bidding close to the expected D-LMPs and awarded more in the market. %It should be noted that most  convergence bidders had a fix strategy regardless of location and time. \color{black} %

\subsection{Comparison Among   Bidding Strategies}
Market participants that used CB Strategies 1, 2, and 3 earned an average monthly profit of about \$104K, \$16K, and \$33K, respectively. CB Strategy 1 has the highest profit while it is associated with the difficulties and risks of forecasting both D-LMPs and R-LMPs. CB Strategy 3 has lower profit while it is a risk-averse strategy and does not require accurate price forecasting. Both CB Strategy 1 and 3 have some advantages and disadvantages and can be considered as two \emph{successful} yet \emph{very different} convergence bidding strategies in the California ISO market. Also, CB Strategy 2 which is adopted by  very few market participants, has the lowest profit and does not appear to be a successful strategy.

\section{Conclusions and Future Work} \label{sec:conclusions}
This paper provided a data-driven analysis of the CBs in the California ISO energy market. About 21\% of the market nodes hosted any CB and a total of 101 convergence bidders  participating in the market. A total of 13 market participants dominated the market and submitted over 75\% of all the CBs. The performance of these dominant market players was analyzed and it was shown that increasing the number of submitted bids or the number of locations does not necessarily result in increasing profit. Next, in order to analyze the strategy of convergence bidders, two quantitative features were introduced. Clear distinctions were observed among market participants in terms of both features. This resulted in introducing three different bidding strategies. It was shown that, while CB Strategies 1 and 3 have some advantages and disadvantages, both can be considered as two successful yet very different convergence bidding strategies in the California ISO market. CB Strategy 2 did not have a successful performance and was adopted by only a few market participants. The results in this paper shed light on the reality of convergence bidding strategies in practice. 
%\color{black}

The study in this paper can be extended in several directions. First, based on our analysis of real-world CB strategies, we can investigate how the current bidding strategies in practice may affect positively or negatively on price convergence or price divergence; thus characterizing the type of behavior by CB market participants that may not be aligned with the ISO's objectives of supporting convergence bidding. Second, we can further investigate the convergence bidding strategies of market participants jointly with the physical bidding strategies. It should be noted that some market participants tend to submit both convergence bids and physical bids. Third, we can learn from the existing bidding strategies in the market to develop new bidding strategies for participation in the convergence bidding market. \color{black} Fourth, the identified CB strategies can be studied  in the context of affecting the overall efficiency in the market outcomes and welfare distribution impact. \color{black}

%systematically develop new bidding . Second, we can expand our analysis to physical bids for different products such as energy and ancillary services to understand the real-world bidding strategies. Third, we can propose a comprehensive bidding strategy for market participants who own physical assets but would like to make money both with physical bids and convergence bids.

\color{black}

%\vspace{-0.1cm}
\bibliographystyle{IEEEtran}
\bibliography{Ehsan_IEEE.bib}
\end{document}